\documentclass{emulateapj}
\usepackage{apjfonts}

\usepackage{graphics,graphicx,rotating}
\usepackage{natbib}
\usepackage{xspace}
\usepackage{amssymb}
\usepackage{amsmath}
\usepackage{epstopdf}
\usepackage{subfigure}

\shorttitle{Offsets Between the Sunyaev--Zel'dovich Effect and X--ray Peaks}
\shortauthors{Molnar, et al.}

\newcommand{\simless} 
     {\ensuremath{\lower 3pt\hbox{$\rlap{\raise5pt\hbox{$\char'074$}}\mathchar"7218$}}}
\newcommand{\simgreat}
     {\ensuremath{\lower 3pt\hbox{$\rlap{\raise5pt\hbox{$\char'076$}}\mathchar"7218$}}}

\newcommand{\simgt}{\lower.5ex\hbox{$\; \buildrel > \over \sim \;$}}
\newcommand{\simlt}{\lower.5ex\hbox{$\; \buildrel < \over \sim \;$}}

\newcommand{\nop}{{\noindent}}

\newcommand{\LCDM}{{\sc $\Lambda$CDM}}

\newcommand{\XMM}{{\sc XMM-Newton}}

\newcommand{\CLJMAS}{{\sc CL J0152-1347}}
\newcommand{\RXJMASON}{{\sc RX J1347-1145}}
\newcommand{\ATCA}{{\sc ATCA}}
\newcommand{\FLASH}{{\sc FLASH}}
\newcommand{\HE}{hydrostatic equilibrium}

\newcommand{\Degree}{{$^{\circ}$}}
\newcommand{\ASEC}{\ensuremath{\arcsec}}
\newcommand{\AMIN}{\ensuremath{\arcmin}}

\newcommand{\MSUN}{{\ensuremath{\mbox{\rm M}_{\odot}}}}
\newcommand{\MSUNFOUR}{{$10^{\,14}\,$\ensuremath{\mbox{\rm M}_{\odot}}}}
\newcommand{\TMSUNFOUR}{{$\times 10^{\,14}\,$\ensuremath{\mbox{\rm M}_{\odot}}}}

\newcommand{\KMSEC}{{$\rm km \,s^{-1}$}}

\newcommand{\rmsub}[1]{\ensuremath{_{\rm #1}}}
\newcommand{\RVIR}{\ensuremath{R\rmsub{vir}}}

\begin{document}

\title{Merging Galaxy Clusters: \\ Offset Between the Sunyaev--Zel'dovich Effect and X--ray Peaks}

\author{ Sandor. M. Molnar\altaffilmark{1}, Nathan C. Hearn\altaffilmark{2}, and Joachim G. Stadel\altaffilmark{3} }

\altaffiltext{1}{Leung Center for Cosmology and Particle Astrophysics, 
                      National Taiwan University, Taipei 10617, Taiwan, Republic of China; sandor@phys.ntu.edu.tw}
                                            
\altaffiltext{2}{Computational \& Information Systems Laboratory, 
                      National Center for Atmospheric Research, PO Box 3000, Boulder CO, 80305, USA}

\altaffiltext{3}{Institute for Theoretical Physics, University of Zurich, 8057 Zurich, Switzerland}

\begin{abstract}
Galaxy clusters, the most massive collapsed structures, have been routinely used 
to determine cosmological parameters. When using clusters for cosmology, the 
crucial assumption is that they are relaxed. However, subarcminute resolution 
Sunyaev--Zel'dovich (SZ) effect images compared with high resolution X--ray 
images of some clusters show significant offsets between the two peaks. We have 
carried out self-consistent N-body/hydrodynamical simulations of merging galaxy 
clusters using \FLASH\ to study these offsets quantitatively. We have found that
significant displacements result between the SZ and X-ray peaks for large 
relative velocities for all masses used in our simulations as long as the 
impact parameters were about 100--250 kpc. Our results suggest that the SZ 
peak coincides with the peak in the pressure times the line-of-sight 
characteristic length and not the pressure maximum (as it would for clusters 
in equilibrium). The peak in the X--ray emission, as expected, coincides with
the density maximum of the main cluster. As a consequence, the morphology of 
the SZ signal and therefore the offset between the SZ and X-ray peaks change 
with viewing angle. As an application, we compare the morphologies of our 
simulated images to observed SZ and X--ray images and mass surface densities 
derived from weak lensing observations of the merging galaxy cluster CL0152-1357.
We find that a large relative velocity of 4800 km/s is necessary to explain these
observations. We conclude that an analysis of the morphologies of multi-frequency 
observations of merging clusters can be used to put meaningful constraints on the 
initial parameters of the progenitors.
\end{abstract}


% % % % % % % % % % % % % % % % % % % % % % % % % % % % % % % % % % % % % % % % % % % %
\section{Introduction}
\label{S:Intro}

Our most successful cosmological model, the cold dark matter model
dominated by the cosmological constant ($\Lambda$CDM), predicts that 
clusters of galaxies form from the largest positive matter density fluctuations. 
The distribution and evolution of these rare, very large positive density fluctuations are extremely 
sensitive to the underlying cosmological model. Therefore clusters have been extensively used
to put constraints on cosmological models.
In general, we can define two main categories to determine cosmological
parameters using galaxy clusters: individual and statistical methods.

Individual methods use accurate measurements of clusters to derive cosmological 
parameters.
The most extensively used individual method is the SZ--X--ray method, which is 
taking advantage of the different dependence of the SZ signal and the X-ray emission on
the physical parameters of galaxy clusters to derive the distances to them directly.
Measuring the redshift to clusters, this method is making use of the distance--redshift function 
to derive cosmological parameters
(\citealt{Bonaet06ApJ647p25,Schmet04MNRAS352p1413};
for prospects for future surveys see 
\citealt{Moln04APJ601p22,Molnet02PAJ570p1,Haimet01APJ553p545,Hold01APJ560pL111}).
Another individual method, for example, is based on the gas mass fraction--redshift 
function \citep{Alleet08MNRAS383p879,Ettoet09AA501p61}.
In theory, resonant line scattering along with X--ray or SZ imaging 
can also be used to derive distances to clusters \citep{Molnet06ApJ643pL73}.

Statistical methods use the average properties of many clusters to derive
cosmological parameters, such as, for example, 
the cluster mass function \citep{Mantet10MNRAS406p1759,Vikhlet09ApJ692p1060},
the X-ray luminosity and temperature function 
\citep{Popoet10AA514p80,Schuet03AA398p867,Borget01APJ561,AnBa11APH1102.0458}.
Using SZ and X--ray number counts based on future surveys are also a promising possibility 
\citep{MaMo03APJ585,LevSchWhi02APJ577,WeBaKn02PHRL88,Haimet01APJ553p545,Hold01APJ560pL111}.
Recent reviews on using galaxy clusters for cosmology can be found in 
\cite{Alleet11} and \cite{RoBoNo02ARAA40p539}.

For both main methods of using clusters to determine cosmological parameters,
it is crucial to understand the physics of clusters, the distribution of the
different components, and their formation history. 
Our $\Lambda$CDM models predict that massive clusters form by 
merging. Therefore some fraction of clusters must be in a merging state. 
Merging clusters with small mass ratios result in  
enhanced X--ray luminosities and temperatures 
relative to relaxed clusters, as shown by using numerical simulations
\citep{Randet02ApJ577p579,RiTh02MNRAS329,RiSa01ApJ561p621}.
The derived mass for a cluster in this stage will be overestimated, thus 
the derived mass function will be biased.
When using individual methods it is crucial to be able to identify mergers 
and exclude them from our analysis. Statistical methods need to be
corrected for the affect on different statistical properties of clusters.
Assuming a LCDM cosmology, \cite{Wik.et.al2008ApJ680p17} have found that 
the maximum of the Comptonization parameter, $y$,
is increased substantially during the first core passage, resulting a significant
bias of 20\%--40\%  in determining $\Omega_m$ and $\sigma_8$,
while cosmological parameters derived from the integrated Compton-y parameter 
introduce only less or equal to 2\% bias.

Subarcminute resolution Sunyaev--Zel'dovich (SZ) effect images of some
clusters of galaxies compared with high resolution X-ray images show that the positions
of the maxima of the SZ and X--ray signals differ significantly 
\citep{Massardi.et.al2010,Korngut.et.al2010,Mason.et.al2010ApJ716p739,Malu.et.al2010,Rodriguez.et.al2010}.
The observed offsets imply that these clusters are not in dynamical equilibrium, therefore the derived
physical parameters fro them, and the derived cosmological parameters would also be biased.

In this paper we study quantitatively the offsets between the 
SZ and X--ray peaks after the first core passage using 
3-dimensional N-body/hydrodynamical simulations of idealized binary  
merging clusters of galaxies using \FLASH.
We focus on the offsets after first core passage since this phase
is the easiest to analyze both observationally and theoretically.

\smallskip
% % % % % % % % % % % % % % % % % % % % % % % % % % % % % % % % % % % % % % % % % % % %
\section{Offsets Between SZ and X--ray Peaks in Galaxy Clusters}
\label{S:FLASH}

\cite{Massardi.et.al2010} observed \CLJMAS\ using the Australia Telescope Compact Array (ATCA)
at 18 GHz with an angular resolution of $35\arcsec \times 35\arcsec$.
\CLJMAS, at a reshift of 0.83, 
is one of the most massive high redshift clusters known. 
\cite{Massardi.et.al2010} have found that the X-ray center based 
on \XMM\ and their \ATCA\ observations 
are displaced by about 342 kpc ($\approx$45\ASEC).

\cite{Malu.et.al2010} observed the ``bullet cluster '' (1E 0657--56) at a 
resolution of 30\ASEC\ using ATCA. They found two X-ray
peaks at about 1.5\AMIN apart at the 
East (E) and West (W) part of the cluster. The peak at W belongs
to the infalling cluster, the ``bullet''.
The offset between the SZ and X-ray peaks were found to be about 
35\ASEC.

An offset of about 20\ASEC\ was found between the SZ and X-ray peaks in 
\RXJMASON\ by \cite{Korngut.et.al2010} (see also \citealt{Mason.et.al2010ApJ716p739})
using Mustang on GBT with about 10\ASEC--18\ASEC\ resolution  
confirming an offset which had been discovered earlier using the lower
resolution and sensitivity Nobeyama Bolometer Array on the Nobeyama 45-m telescope 
\citep{Komatsu2001PASJ53p57,Kitayamaet2004PASJ56p17}.
Using Mustang, \cite{Korngut.et.al2010} observed two more disturbed clusters
and found a kidney-shaped SZ feature between the two peaks
of the X-ray emission displaced by about 20\ASEC\ in MACS0744, 
and a highly disturbed SZ distribution with multiple peaks with a prominent 
ridge like feature  oriented perpendicular to the line connecting the 
center of the main and secondary total mass distributions in CL1226.
Their preliminary study of disturbed clusters demonstrated the 
importance of high-resolution SZ observations in identifying shocks in
galaxy clusters particularly at high redshifts.

\cite{Rodriguez.et.al2010} found an about 20\ASEC\ displacement between
the X-ray and SZ peaks in A2146 using the Arcminute Microkelvin Imager (AMI).
AMI has a 3\AMIN\ resolution in its most compact configuration, 
which is suitable for extended sources, and an about 30\ASEC\ resolution
in its more extended configuration to remove point sources.

\smallskip
% % % % % % % % % % % % % % % % % % % % % % % % % % % % % % % % % % % % % % % % % % % %
\section{FLASH Simulations of Merging Galaxy Clusters}
\label{S:FLASH}

A number of self--consistent 3-dimensional (3D) binary merger simulations have been 
carried out recently using Lagrangian 
(Ritchie \& Thomas 2002; McCarthy et al. 2007; Poole et al. 2006, 2007),
and Eulerian codes (Ricker \& Sarazin 2001; ZuHone 2010) to study the
effects of different mass ratios, impact parameters and cluster models on
X-ray emission, SZ signal and mass surface density.
Binary merger simulations were also carried out to compare numerical
simulations with X-ray and SZ observations of individual clusters
(for the ``bullet cluster'', Cl 1E0657-56, 
Springel \& Farrar 2007 and Mastropietro \& Burkert 2008; and for 
Cl 0024+17, ZuHone et al. 2009a,b).
Binary merger simulations using self consistent N--body 
smoothed particle hydrodynamics (SPH) and Eulerian (FLASH) have been carried 
out by \cite{Mitcet09MNRAS395p180} to compare the results for mixing/turbulence
for these two different methods. \cite{Mitcet09MNRAS395p180} showed that 
SPH codes suppress turbulence, while adaptive mesh refinement (AMR)
codes treat them more realistically
(see also \citealt{Agertzet2007MNRAS380p963}).

Since turbulence is important in cluster mergers, we choose to use a publicly
available parallel Eulerian parallel code, \FLASH, 
developed at the Center for Astrophysical Thermonuclear Flashes
at the University of Chicago \citep{Fryxell2000ApJS131p273}.
\FLASH\ is using the Piecewise-Parabolic Method (PPM) of Colella \& Woodward (1984) 
to solve the equations of hydrodynamics, and a particle-mesh method to solve for the 
gravitational forces between particles in the $N$-body module. 
The gravitational potential is calculated using a 
multigrid solver (Ricker 2008). \FLASH\ uses adaptive mesh refinement (AMR) with a 
tree-based data structure allowing recursive grid refinements on a cell-by-cell basis
on a Cartesian grid.

Our simulations achieved a 12.7 kpc resolution at the cluster centers and the merger shocks. 
Our box size, 13.3 Mpc on a side, was large enough, therefore there was no need for 
corrections for mass loss.

\subsection{Initial Conditions}
\label{SS:INIT}

We assumed spherical cluster models with a cut off of the distribution
of the dark matter and gas at the virial radius, \RVIR\ \citep{BrNo98APJ495}.
We used an NFW model \citep{NFW1997ApJ490p493} for the dark matter density,

\begin{equation}  \label{E:RHODM}
      \rho_{DM} (r) =  { \rho_s  \over x (1 + x)^2}
,
\end{equation}
% Equation~\ref{E:RHODM}
\nop
where $x = r/r_s$, and $\rho_s$, $r_s = r_{vir}/c_{vir}$ are scaling parameters for the density and radius, 
$c_{vir}$ is the concentration parameter, and $r \le R_{\rm vir}$.
The gas distribution was assumed to be a truncated non-isothermal $\beta$ model, 

\begin{equation}  \label{E:RHODM}
      \rho (r) =  { \rho_0  \over (1 + y^2)^{3 \beta /2} }
,
\end{equation}
% Equation~\ref{E:RHODM}
\nop 
where $y = r/r_{core}$, and $\rho_0$, $r_{core}$ are the central density and gas scale radius,
and $r \le R_{\rm vir}$.
The temperature of the gas was determined form the equation of \HE\ via numerical integration.
The equation of state for the gas was an ideal gas equation of state with $\gamma = 5/3$, 
and the mean atomic mass was $\mu = 0.592$. We adopted $r_{core} = 0.12$ \RVIR, and $\beta=1$,
which are consistent with our analysis of numerical simulations \citep{Molnet10ApJ723p1272}.
In our simulations presented in this paper, 
we used 5 and 8 for $c_{vir}$, for the main and infalling (sub)cluster 
following the trend that less massive clusters are more concentrated. 
We have also run some simulations
with different concentration parameters to check their effect. 
Assuming $M_{tot}$ and the gas mass fraction of 0.14 
we derive $R_{vir}$, $\rho_s$, $r_s$, $\rho_0$. 
We treat the small fraction of baryonic matter in galaxies along with the dark matter
since they can be assumed to be collisionless for our purposes. 
Therefore our dark matter particles also represent baryonic matter in galaxies.
The number of dark matter particles at each cell 
was determined by the density and the total particle number (5 million).

% % % % % % % % % % % % % % % % % % % % % % % % % % % % % % % % % % % % % % % % % % % %
%  
%   TABLE 1
% 
% % % % % % % % % % % % % % % % % % % % % % % % % % % % % % % % % % % % % % % % % % % %
 \begin{table}
 \centering
 \caption{Initial Parameters}
 \begin{tabular}{|c|c|c|c|c|} \hline
    ID                              &    $M_1$   &   $M_2$  &    $V$    &    $P$     \\ \hline
    RM1V48p00            &       2.1      &   1.0        &    4800  &        0     \\ \hline
    RM1V48p10            &       2.1      &   1.0        &    4800  &    100     \\ \hline
    RM1V48p15            &       2.1      &   1.0        &    4800  &    150     \\ \hline
    RM1V48p20            &       2.1      &   1.0        &    4800  &    200     \\ \hline
    RM1V48p25            &       2.1      &   1.0        &    4800  &    250     \\ \hline
    RM1V48p35            &       2.1      &   1.0        &    4800  &    350     \\ \hline
    RM1V45p15            &       2.1      &   1.0        &    4800  &    350     \\ \hline
    RM1V40p15            &       2.1      &   1.0        &    4000  &    150     \\ \hline
    RM1V35p15            &       2.1      &   1.0        &    3500  &    150     \\ \hline
    RM1V30p15            &       2.1      &   1.0        &    3000  &    150     \\ \hline
    RM1p6V48P15        &       2.1      &   1.6        &    4800  &    150     \\ \hline
    RM1p3V48P15        &       2.1      &   1.3        &    4800  &    150     \\ \hline
    RM0p7V48P15        &       2.1      &   0.7        &    4800  &    150     \\ \hline
 \end{tabular}
 \label{T:TABLE1} 
 \tablecomments{See text for other parameters}
\end{table}
%Table~\ref{T:TABLE1}

The velocities of the dark matter particles were determined by 
sampling a Maxwellian distribution (local Maxwellian approximation)
with the velocity dispersion, $\sigma_r$ derived from the Jeans equation 
\citep{LokasMamon2001MNRAS.321p155}.
Assuming isotropic velocity dispersion 
(the angular and radial components are equal: $\sigma_\theta = \sigma_r$), 
the Jeans equation becomes:

\begin{equation}  \label{E:JEANS}
     {1 \over \rho} {d \over dr} \bigl[ \rho \sigma_r^2 \bigr] = - {d \Phi \over dr} ,
\end{equation}
% Equation~\ref{E:JEANS}
\nop 
where $\Phi(r)$ is the gravitational potential, which, in our case of
an NFW distribution, becomes

\begin{equation}  \label{E:FI}
    \Phi (r) = - V_v^2 \, g(c_{vir}) \,  {\ln q \over  x }
,
\end{equation}
% Equation~\ref{E:GC}
where the circular velocity is $V_v^2 = G M_{vir} / R_{vir}$,

\begin{equation}  \label{E:GC}
    g(c_{vir})  =  \bigl[ \ln (1 +  c_{vir} )  -  c_{vir} / ( 1 +  c_{vir} ) \bigr]^{-1}
,
\end{equation}
% Equation~\ref{E:GC}

\nop
and $q = 1 + x \, c_{vir}$.
Thus, from Equation~\ref{E:JEANS}, we obtain the velocity dispersion:

\begin{equation}  \label{E:SIGMA}
    \sigma_r^2 (r) = V_v^2 \, g(c_{vir}) \, x \, q^2 \int_x^\infty  
                           \Biggl[ {  \ln q \over  x^3 q^2 } - {  c_{vir} \over x^2 q^3 } \Biggr]  dx
,
\end{equation}
% Equation~\ref{E:SIGMA}

\nop 
(see \citealt{LokasMamon2001MNRAS.321p155} for more details on the properties of
NFW profiles).
The direction of the velocities were assumed to be isotropic. 
Running a control simulation with an isolated cluster, we found that 
this local Maxwellian approximation will relax the density distribution close to the 
assumed NFW model (as expected).
Therefore we conclude, that this approximation is adequate for our purposes,
since we are interested in the effect of merging on the gas distribution and not in the 
details of the changes in the dark matter distribution.

% % % % % % % % % % % % % % % % % % % % % % % % % % % % % % % % % % % % % % % % % % % %
%  
%  FIGURE 1
% 
% % % % % % % % % % % % % % % % % % % % % % % % % % % % % % % % % % % % % % % % % % % %
\begin{figure*}
\centerline{
\includegraphics[width=.95\textwidth]{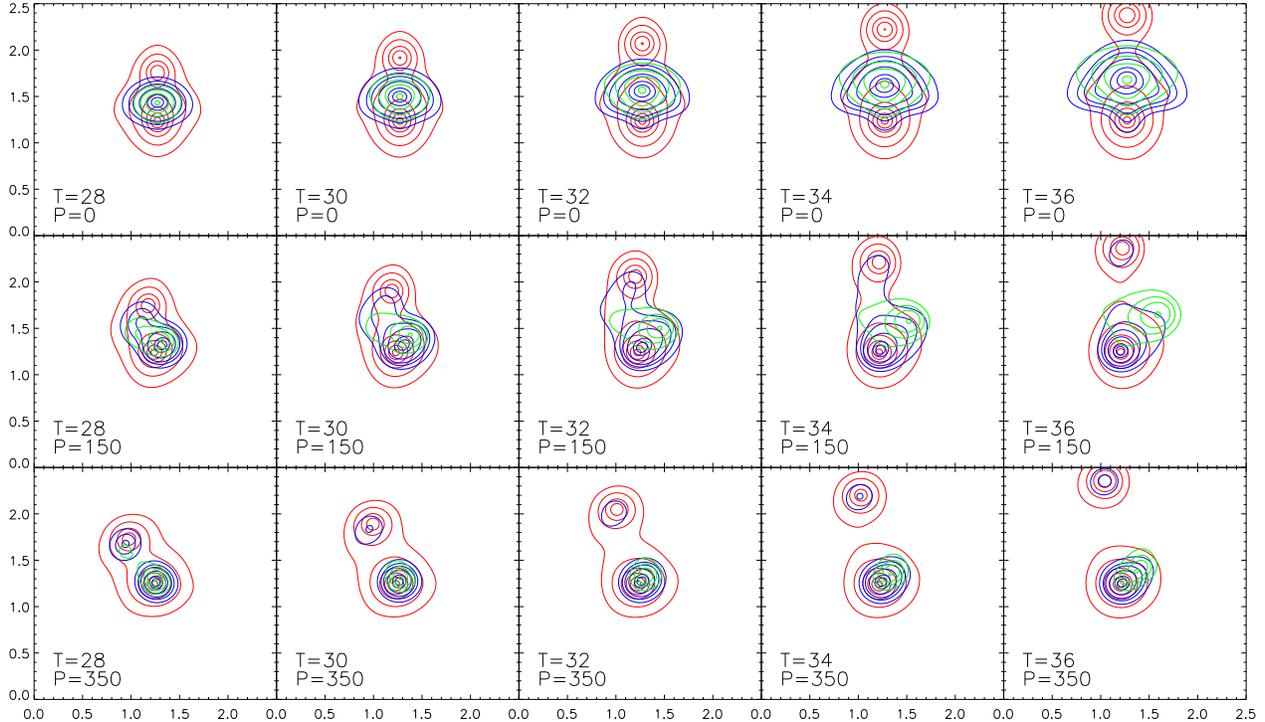}
%FIGDELP0527.eps}
}
\caption{
 Snapshots of merging galaxy clusters (infall velocity is upward)
 after the first core passage at time, $T$, in output time units (left to right), 
 assuming different impact parameters, $P$ = 0, 150, and 350 kpc (from top to bottom).
 All other parameters held fixed 
 (runs RM1V48p00, RM1V48p15, and RM1V48p35; see Table 1 for details).
 We show contour plots of projected mass surface density, X-ray emission 
 and SZ contours (red, blue and green lines). 
 The contour levels are chosen arbitrarily to make it easier to compare morphologies
 and avoid clutter.
 The collision is in the plane of the sky (the projection axis is in the line of sight).
 The images are 2.5 Mpc $\times$ 2.5 Mpc.
\label{F:DELP4}
}
\end{figure*} % Figure~\ref{F:DELP4}

% % % % % % % % % % % % % % % % % % % % % % % % % % % % % % % % % % % % % % % % % % % %
%  
%  FIGURE 2
% 
% % % % % % % % % % % % % % % % % % % % % % % % % % % % % % % % % % % % % % % % % % % %
\begin{figure*}
\centerline{
\includegraphics[width=.95\textwidth]{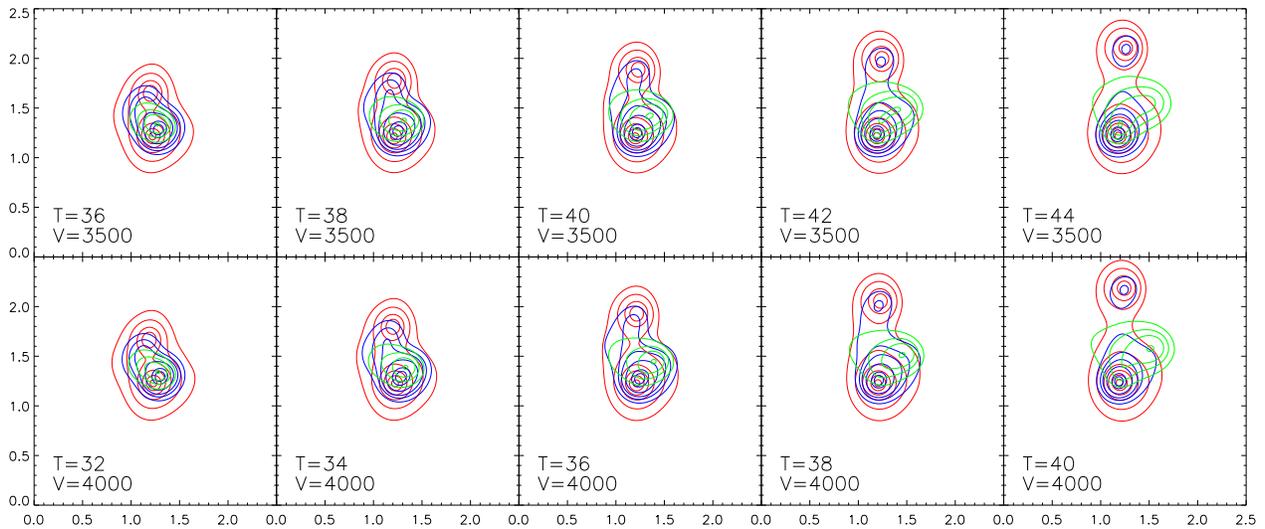}
%FIGDELV0530.eps}
}
\caption{
 Same as Figure~\ref{F:DELP4} but 
 assuming different initial relative velocities, 3500 and 4000, 
 in unit of km s$^{-1}$ (from top to bottom),
 all other parameters held fixed 
 (runs RM1V35p15 and RM1V40p15; see Table 1 for details of the simulations).
 \label{F:DELV4}
}
\end{figure*} % Figure~\ref{F:DELV4}

\subsection{FLASH 3D Simulations}
\label{SS:Simulations}

We have run a set of simulations of galaxy cluster mergers systematically changing
the initial mass ratios, impact parameters, and relative velocities.
In Table~\ref{T:TABLE1} 
we show the initial parameters for those runs we discuss in detail in our paper.
In this table the first column is the identification number for our runs using 
the following convention: the numbers after M, V and P represent the
mass of the second cluster in units of $10^{\,14}\,$~\MSUN, the initial 
relative velocity in units of 100 \KMSEC, and the impact parameter in units of 10 kpc.
The mass of the main cluster was assumed to be 2.1 \TMSUNFOUR, 
The concentration parameters were assumed to be 5 and 8 for the main and the infalling
cluster.
We have also run a few simulations with different gas mass fractions, concentration
parameters and mass rations, and one simulation with higher resolution.

\smallskip
% % % % % % % % % % % % % % % % % % % % % % % % % % % % % % % % % % % % % % % % % % % %
\section{SZ, X-ray and Surface Mass Density Images}
\label{S:Observables}

After the simulation finished, we generated SZ, X--ray and surface mass density, 
$\Sigma$, images assuming different viewing angles 
(expressed as rotation angles out of the plane of the sky) and phases 
(time elapsed after first core passage) of the collisions.

Since relativistic corrections are important in the high temperature shocked gas
($T_g \simgreat$ 30 keV), we generated SZ and X--ray images using relativistic corrections.
We calculated the SZ surface brightness using 
\begin{equation}  \label{E:XRAY}
    I_{SZ}\, (x,y) \propto  \int_{\ell_1}^{\ell_2}
                      \rho_g \, T_g \bigl[ g(\nu) +  \Sigma_{n=1}^{n=4} Y_n \, \Theta^n \bigr] \, d \ell
,
\end{equation}
% Equation~\ref{E:XRAY}
where $\Theta = k_B T_g / (m_e c^2)$ and $g(\nu) = {\text{coth}}( x_\nu/2) - 4$, 
$x_\nu = h_P\, \nu / ( k_B T_{cmb})$, and $\nu$ is the frequency, $T_{cmb}$ is
the temperature of the cosmic microwave background, 
$h_P$, $k_B$, $c$ and $m_e$, 
are the Planck and Boltzmann constants, the speed of light and the electron mass.
The relativistic corrections, $Y_{1,2,3,4}$, were taken from \cite{itohet98ApJ502}.

We generated the X--ray images using \cite{RybLig79}'s expression for the relativistic X-ray 
thermal bremsstrahlung,

\begin{equation}  \label{E:XRAY}
    I_X (x,y) \propto  \int_{\ell_1}^{\ell_2}
                           \rho_g^2 \, T_g^{1/2} g_{ff} \, (1 + 4.4 \times 10^{-10} T_g ) \; d \ell
,
\end{equation}
% Equation~\ref{E:XRAY}
where the Gaunt factor, $g_{ff} = (2 \sqrt(3) / \pi )  [ 1 + 0.79 \,(4.95 \times 10^5/ T_g) ] $
(see also \citealt{HugBir98ApJ501p1}).

We integrated the total (dark matter and gas) density along the line of sight (LOS)
from different viewing angles to obtain the mass surface density images at 
position, $x$ and $y$:

\begin{equation}  \label{E:SIGMA}
    \Sigma_X (x,y) \propto  \int_{\ell_1}^{\ell_2} \, ( \rho_d + \rho_g )  \; d \ell
,
\end{equation}
% Equation~\ref{E:SIGMA}
where $\ell$ is the spatial coordinate in the LOS.

% % % % % % % % % % % % % % % % % % % % % % % % % % % % % % % % % % % % % % % % % % % %
%  
%  FIGURE 3
% 
% % % % % % % % % % % % % % % % % % % % % % % % % % % % % % % % % % % % % % % % % % % %
\begin{figure*}
\centerline{
\includegraphics[width=.95\textwidth]{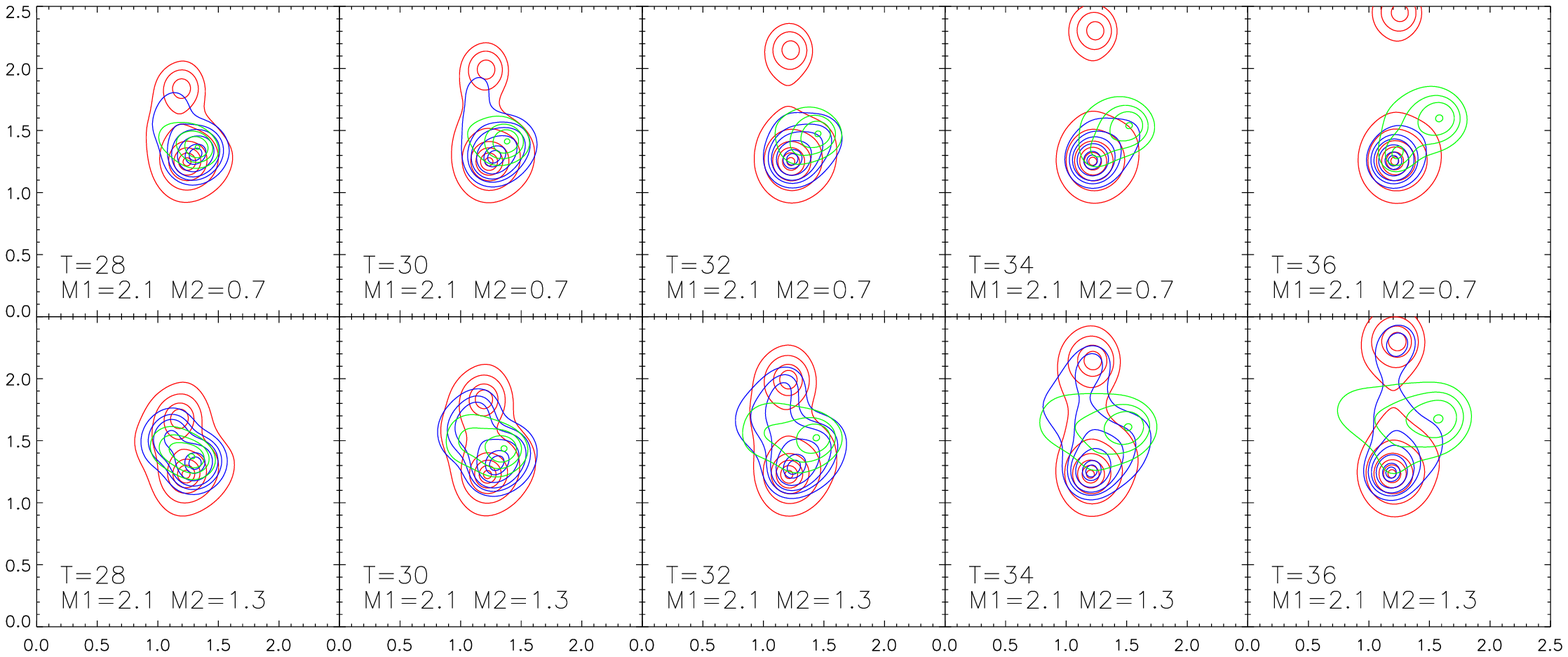}
%FIGDELM0530.eps}
}
\caption{
 Same as Figure~\ref{F:DELP4} but 
 assuming different mass rations: 
 the mass of the main cluster was fixed at M$_1$ = 2.1, and the infalling cluster 
 with masses: M$_2$ = 0.7 and 1.3 in unit of 10$^{14}$ \MSUN\ (from top to bottom),
 all other parameters held fixed (see text for details).
\label{F:DELM4}
}
\end{figure*} % Figure~\ref{F:DELM4}

% % % % % % % % % % % % % % % % % % % % % % % % % % % % % % % % % % % % % % % % % % % %
%  
%  FIGURE 4
% 
% % % % % % % % % % % % % % % % % % % % % % % % % % % % % % % % % % % % % % % % % % % %
\begin{figure*}
\centerline{
\includegraphics[width=.95\textwidth]{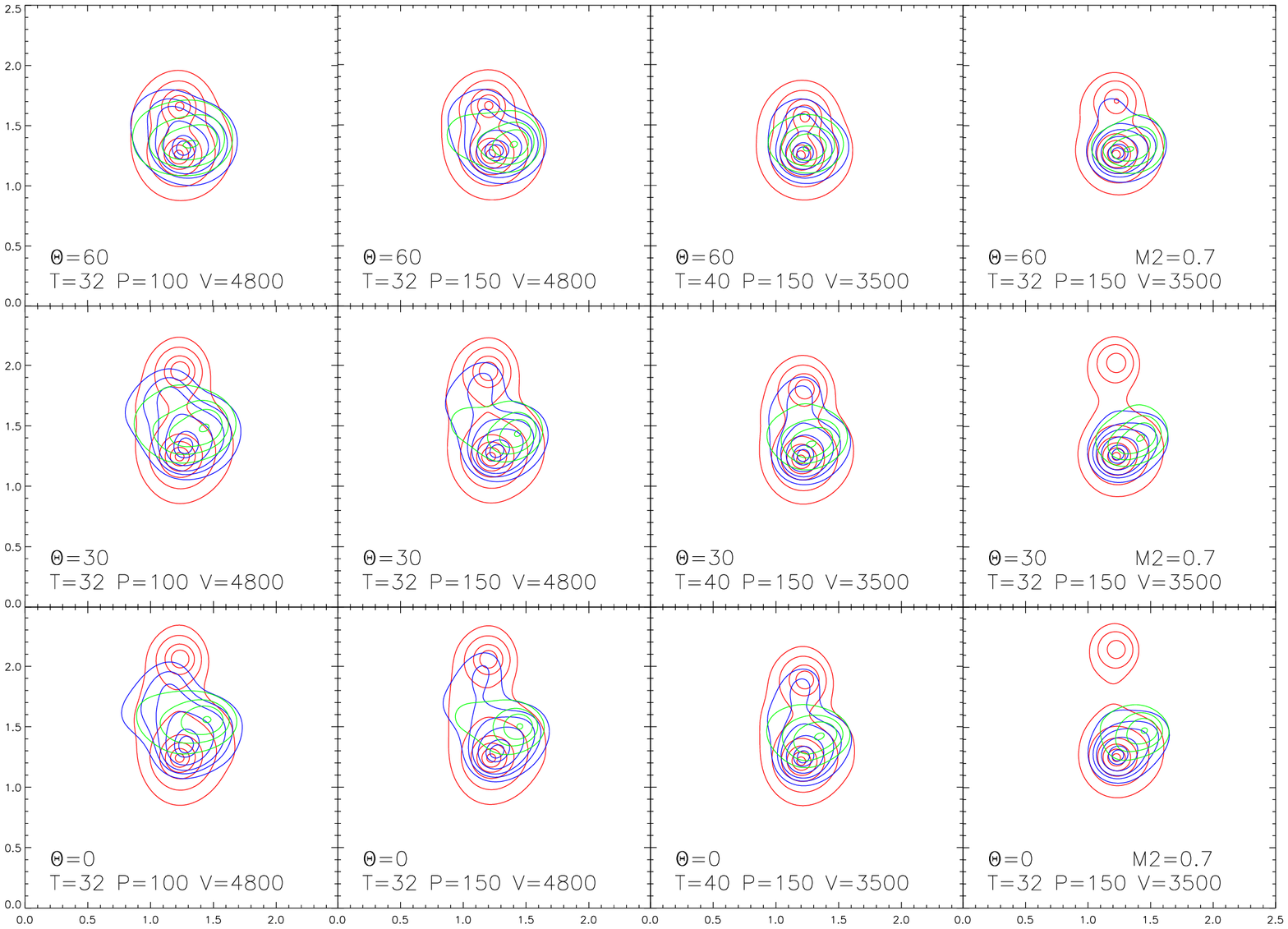}
}
\caption{
 Snap hots of mass surface density, SZ effect and X--ray images of merging clusters of galaxies
 as a function of different rotation angles, 
 $\Theta$ = 0\Degree, 30\Degree, 60\Degree (bottom to top), away from the main plane 
 of the collision with axis perpendicular to the direction of the initial infalling velocity 
 for different initial velocities. The color code is the same as in Figure~\ref{F:DELP4}.
 The infall velocity points upward. The initial masses are: 
 $M_1 = 2.1$ \TMSUNFOUR, $M_2= 1.0$ \TMSUNFOUR, except for the right column.
 From left to right: $P = 100, 150, 150, 150$ kpc, $V = 4800, 4800, 3500, 3500$ in km s$^{-1}$.
 All other parameters help fixed (see text for details).
  The images are 2.5 Mpc $\times$ 2.5 Mpc.
\label{F:THETA3V}
}
\end{figure*} % Figure~\refF:THETA3V}

\smallskip
% % % % % % % % % % % % % % % % % % % % % % % % % % % % % % % % % % % % % % % % % % % %
\section{Results: Offsets Between the X-ray and SZ Centers}
\label{S:Results}

In this paragraph we discuss general aspects of our results.
We have run simulations with different initial galaxy cluster masses and found similar dependence of 
the offsets between X-ray and SZ centers on different impact parameters and relative velocities.
Thus we choose to show our results with fixed $M_1 = 2.1 \times 10^{14}$ \MSUN, since these  
give the best match with the observed morphology of CL0152-1357, which we will discuss in the next section.
In Figure~\ref{F:DELP4} we show 
snapshots of merging galaxy clusters after the first core passage
assuming different impact parameters, $P = 0$, 150 and 350 kpc,
all other parameters of the simulations were held fixed 
(runs RM0p7V48P15 and RM1p3V48P15, see Table 1 for details of the simulations).
The contour plots centered on the main cluster show the mass surface density, $\Sigma$, 
the X-ray emission and SZ contours (red, blue and green lines) projected to the main 
plane of the collision  containing the two mass centers and the relative velocity vector
(i.e., the collision is in the plane of the sky).
$T$ represents the elapsed time in output file time units.
The images are 2.5 Mpc $\times$ 2.5 Mpc.

The most obvious feature we recognize looking at Figure~\ref{F:DELP4}
is the different behavior of the infalling cluster gas and the most 
massive component, the dark matter. As we expected, the dark matter in the
infalling cluster, due to its collisionless nature, simply passes through the 
main cluster smoothly, while the gas gets shocked because of the collision with
the gas of the main cluster, and it is slowed down due to ram pressure.
As a consequence, the gas of the infalling cluster will be out of equilibrium, and 
it gets displaced relative to the dark matter, it falls behind the center of the 
infalling cluster dark matter.

The effect of the ram pressure as a function of the impact parameter
can be seen clearly.  At larger impact parameters the core of the infalling cluster
goes through less dense regions of the main cluster and, since the ram pressure
is proportional to density $\times$ velocity$^2$, the infalling cluster retains
more and more of its gas. The displacement between the SZ and X-ray peaks
can be seen in all cases with non-zero impact parameter.

% % % % % % % % % % % % % % % % % % % % % % % % % % % % % % % % % % % % % % % % % % % %
%  
%  FIGURE 5
% 
% % % % % % % % % % % % % % % % % % % % % % % % % % % % % % % % % % % % % % % % % % % %
\begin{figure}
\centerline{
\includegraphics[width=.48\textwidth]{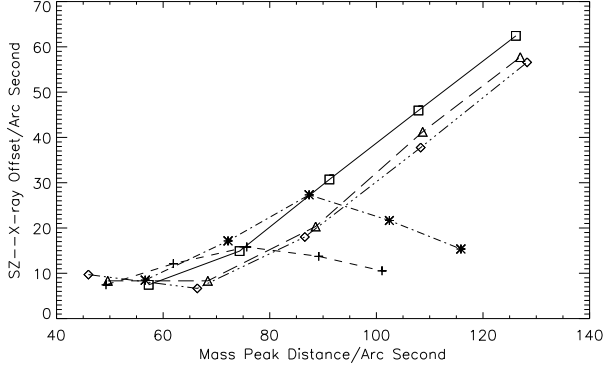}
}
\caption{
 SZ--X--ray offsets as a function of the distance between the dark matter mass
 centers of the two merging clusters for different initial relative velocities: 
 3000, 3500, 4000, 4500, and 4800 \KMSEC\
 (plus signs, stars, squares, triangles and diamonds connected by 
 short dashed, dot-dashed, solid, long dashed, dot-dot-dot-dashed lines).
\label{F:OFFSETV}
}
\end{figure} % Figure~\ref{F:OFFSETV}

% % % % % % % % % % % % % % % % % % % % % % % % % % % % % % % % % % % % % % % % % % % %
%  
%  FIGURE 6
% 
% % % % % % % % % % % % % % % % % % % % % % % % % % % % % % % % % % % % % % % % % % % %
\begin{figure}
\centerline{
\includegraphics[width=.48\textwidth]{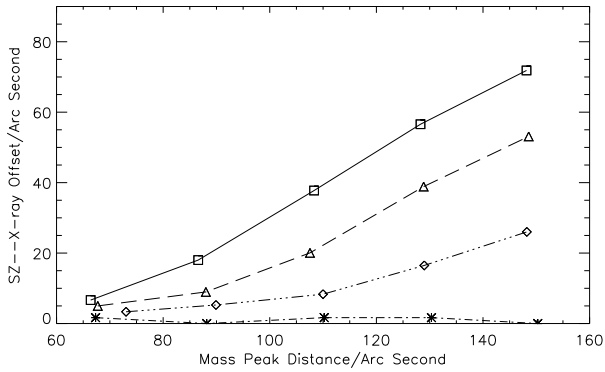}
}
\caption{
 Same as Figure~\ref{F:OFFSETV} but for different impact parameters: P = 0, 150, 250, 350 kpc
 (stars, squares, triangles and diamonds connected by 
 dot-dashed, solid, long dashed, dot-dot-dot-dashed lines).
 All other parameters help fixed (see Table 1 for details of the simulations).
\label{F:OFFSETP}
}
\end{figure} % Figure~\ref{F:OFFSETP}

We can see how the displacement depends on the initial relative velocity 
by studying Figure~\ref{F:DELV4}.
In this figure we show results for $V =$ 3500 \KMSEC\ and 4000 \KMSEC, 
all other parameters held fixed
(runs RM1V35p15 and RM1V40p15; see Table 1 for details of the simulations).
Results from simulations with the same initial conditions but with relative velocities
of $V =$ 4800 \KMSEC\ are shown in the second row of Figure~\ref{F:DELP4}.
Again, we can see that assuming larger relative velocities, the infalling cluster
can retain less of its gas due to ram pressure stripping.
Also, larger velocities result larger displacements between the SZ and
X--ray peaks. However, it seems that if the velocity is large enough, 
\simgt 3500 \KMSEC, we obtain large displacements, \simgt 200--300 kpc
(the largest offset is at $T = 40$ for this relative velocity).

In Figure~\ref{F:DELM4} we show snapshots of the collisions as a function
of different masses of the subcluster.
The mass of the main cluster was fixed at $M_1 = 2.1 \times 10^{14}$ \MSUN, 
and the masses of the infalling cluster were $M_2$ = 0.7 and 1.3 $\times 10^{14}$ \MSUN\
(for $M_2$ = 1.0 see the second row of Figure~\ref{F:DELP4}, for which the other 
 initial parameters are the same).
Again, all other parameters held fixed (see Table 1 for details).
In this case we see the result of two opposing effects: 
the ram pressure is striping the gas of the infalling cluster, and
the gravity is trying to keep the gas, the relative velocities were held fixed.
Assuming $M_2 = 0.7 \times 10^{14}$ \MSUN, the infalling cluster looses all
its gas, and with increasing mass, the subcluster retains more and more
of its gas ($M_2$ = 1, and 1.3 $\times 10^{14}$ \MSUN).

In Figure~\ref{F:THETA3V}, we 
show the projections rotated by angles $\Theta$ = 0\Degree, 30\Degree, 60\Degree, 
away from the main plane of the collision with axis perpendicular to the direction of the 
initial infalling velocity (from bottom to top, 
runs RM1V48p10, RM1V48p15, RM1V35p15 and RM0p7V48P15).
The infall velocity points upward. The initial masses are: 
$M_1 = 2.1$ \TMSUNFOUR, $M_2= 1.0$ \TMSUNFOUR, except for the
 right column for which $M_2= 0.7$ \TMSUNFOUR.
 From left to right: $P = 100, 150, 150, 150$ kpc, $V = 4800, 4800, 3500, 3500$ in km s$^{-1}$.
 All other parameters help fixed (see Table 1 for details).
It is interesting to note the significant changes in SZ morphology, and, as a consequence,
the offset between SZ and X--ray peaks:
the offset gets smaller with larger rotation angle for all cases shown.

% % % % % % % % % % % % % % % % % % % % % % % % % % % % % % % % % % % % % % % % % % % %
%  
%  FIGURE 7
% 
% % % % % % % % % % % % % % % % % % % % % % % % % % % % % % % % % % % % % % % % % % % %
\begin{figure}
\centerline{
\includegraphics[width=.48\textwidth]{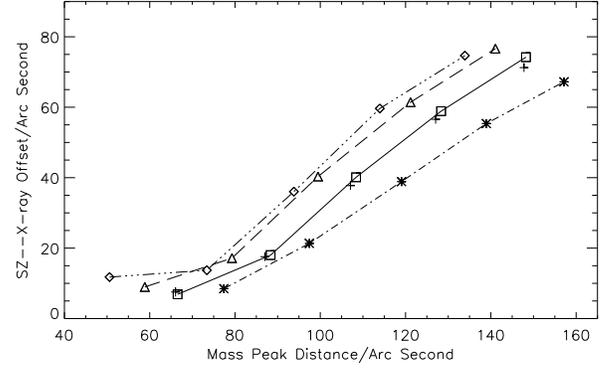}
}
\caption{
 Same as Figure~\ref{F:OFFSETP} but for different initial mass ratios: M$_1$ = 2.1,
 fixed, M$_2$ = 0.7,1.0,1.3,1.6 in units of \MSUNFOUR\
 (stars, squares, triangles and diamonds connected by 
 short dashed, dot-dashed, solid, long dashed, dot-dot-dot-dashed lines).
 The plus signs represent results from a higher resolution run 
 with M$_1 = 2.1 \times$\MSUNFOUR\ and M$_2 = 1 \times$\MSUNFOUR.
\label{F:OFFSETM}
}
\end{figure} % Figure~\ref{F:OFFSETM}

% % % % % % % % % % % % % % % % % % % % % % % % % % % % % % % % % % % % % % % % % % % %
%  
%  FIGURE 8
% 
% % % % % % % % % % % % % % % % % % % % % % % % % % % % % % % % % % % % % % % % % % % %
\begin{figure}
\centerline{
\includegraphics[width=.48\textwidth]{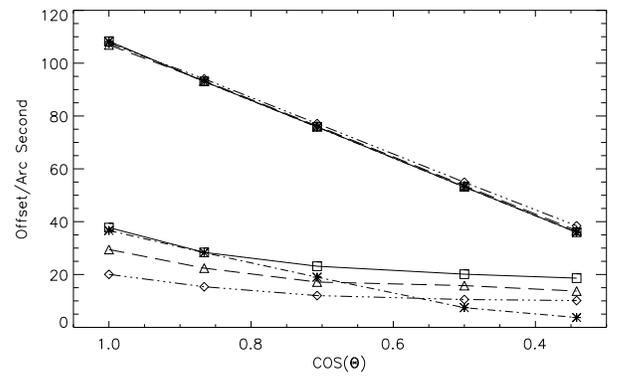}
}
\caption{
 Offsets between the two mass centers and the SZ and X--ray peaks (upper and
 lower set of curves) as a function of the cosine of the rotation angle
 away from the main plane 
 of the collision with axis perpendicular to the direction of the initial infalling velocity 
 for different impact parameters: 100, 150, 200, 250 kpc 
 (stars, squares, triangles and diamonds connected by dot-dashed, solid, dashed, 
 dot-dot-dot dashed lines).
\label{F:THETAP}
}
\end{figure} % Figure~\ref{F:THETAP}

We show our results for the offsets between the SZ and X-ray peaks 
for different initial velocities, impact parameters, and masses 
in Figure~\ref{F:OFFSETV}
(runs RM1V30p15, RM1V35p15, RM1V40p15, RM1V45p15, and RM1V48p15), 
Figure~\ref{F:OFFSETP}
(runs RM1V48p00, RM1V48p15, RM1V48p25, and RM1V48p35), 
and Figure~\ref{F:OFFSETM}
(runs RM0p7V48P15, RM1V48p15, RM1p3V48P15, and RM1p6V48P15).

A qualitative analysis of the offset between the SZ and X--ray peaks
shows that 40\ASEC\ or larger distances between the peaks can 
be produced as long as the relative velocities are at least
4000 \KMSEC\ (see Figure~\ref{F:OFFSETV}).
At large initial relative velocities (V = 4800 \KMSEC) runs with 
impact parameters between 100 and 250 kpc, or, with fixed P = 150 kpc, 
all masses with fixed M$_1$ = 2.1\TMSUNFOUR, and 
M$_2$ = 0.7--1.6\TMSUNFOUR, the offset is larger than 40\ASEC\
(see Figures~\ref{F:OFFSETP} and \ref{F:OFFSETM}).

We have carried our a run using higher resolution to check if our resolution
is high enough. The offsets for this run are included in our  Figure~\ref{F:OFFSETM} 
(plus signs). In this figure, the squares represent offsets based on a run
we used in all simulations. The two offsets are almost identical, thus we 
conclude that our resolution is sufficient for our purpose.

In Figure~\ref{F:THETAP} we illustrate how the offsets between the SZ and X--ray 
change with rotation angle (out of the plane of the collision), 
$\Theta$, for different impact parameters, P = 100, 150, 200, 250 kpc
(runs RM1V48p10, RM1V48p15, RM1V48p20 and RM1V48p25).
The upper lines represent the distances between the 
two mass peaks, the lower lines show the offsets between the SZ and X--ray
peaks. 
Assuming $\Theta = 0$, we find large distances between the projected
mass centers, about 110\ASEC, when the largest
offset between the SZ and X--ray peaks are about 40\ASEC.
At each $\Theta$, the largest offset we obtain is for $P =$ 150 kpc,
$M_1 = 2.1$ \TMSUNFOUR, $M_2= 1.0$ \TMSUNFOUR, and 
$V =$ 4800 \KMSEC. 
This is the largest offset for all masses we considered here.

We conclude that large relative velocities of $V$ \simgt 4000 \KMSEC\
with different impact parameters close to the core radius,
and masses can easily produce 
40\ASEC, or larger offset between SZ and X--ray peaks.

\smallskip
% % % % % % % % % % % % % % % % % % % % % % % % % % % % % % % % % % % % % % % % % % % %
\section{Application to Galaxy Cluster CL0152-1357}
\label{S:CL0152}

In this section we use our results from \FLASH\ simulations to interpret
multi-wavelength observations of the galaxy cluster CL0152-1357.
We use the morphology of the X-ray, SZ and $\Sigma$ images, 
and the offsets between the SZ and X--ray peaks as well as
the distances between the mass peaks of the two components
to constrain the initial conditions of the collision.
We also discuss the implications of our results for CL0152-1357 
as a test for \LCDM\ and the effect of SZ resolution on the interpretation 
of galaxy clusters.

% % % % % % % % % % % % % % % % % % % % % % % % % % % % % % % % % % % % % % % % % % % %
%  
%  FIGURE 9
% 
% % % % % % % % % % % % % % % % % % % % % % % % % % % % % % % % % % % % % % % % % % % %
\begin{figure}
\centerline{
\includegraphics[width=.43\textwidth]{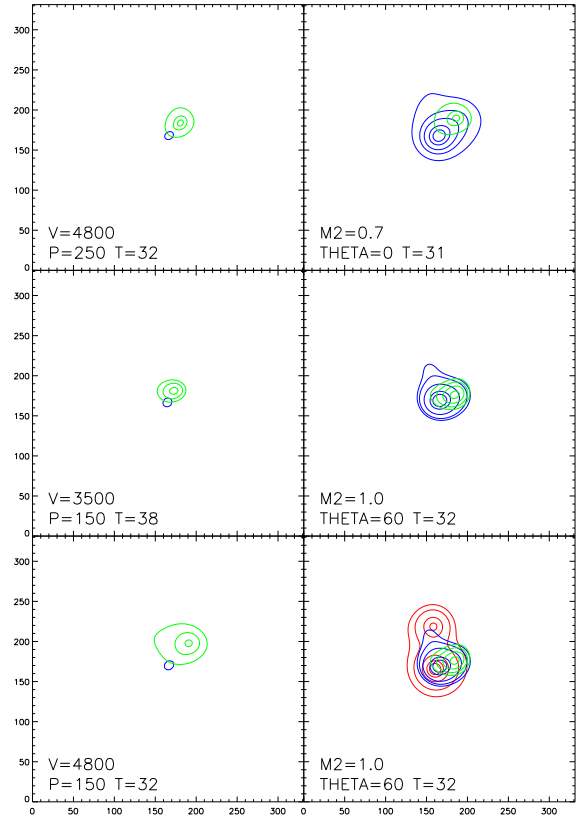}
}
\caption{
Contours of SZ, X-ray and mass surface density (green, blue, red colors)
for different models. Left column: similar SZ morphology to the
observed (to aid the eye, we included the center of the X-ray emission)
with different initial velocities, 3500, 4800 in units of \KMSEC, and 
impact parameters (150 kpc and 250 kpc). Right column: top
2 panels: SZ and X-ray morphology similar to the observed, 
bottom panel: SZ, X-ray and mass surface density 
morphology similar to the observed
(60\Degree\ rotation away form the plane of the collision).
The contour levels are chosen arbitrarily to make an easy
comparison to the observed morphology.
 The images are $350\,\arcsec \times 350\,\arcsec$.
\label{F:MULTI}
}
\end{figure} % Figure~\ref{F:MULTI}

\smallskip
% % % % % % % % % % % % % % % % % % % % % % % % % % % % % % % % % % % % % % % % % % % %
\subsection{Constraining Initial Conditions}
\label{SS:INITIAL}

The ATC SZ image of CL0152-1357 seems to be nearly circularly 
symmetric (see Figure 1 of \citealt{Massardi.et.al2010}).
Searching for similar SZ morphology in our projected images we find candidates 
with different impact parameters and initial relative velocities
(see green contours in the left column of Figure~\ref{F:MULTI},
results for runs RM1V48p25, RM1V35p15 and RM1V48p15), also
with different initial masses (green contours in first panel in Figure~\ref{F:MULTI},
run RM0p7V48P15 with 
$M_1$ = 2.1 \TMSUNFOUR\ and $M_1$ = 0.7 \TMSUNFOUR\
(see Table 1 for other parameters).
We assumed a 35\ASEC\ resolution for SZ observations (ATCA),
and choose the contours to reflect the signal observed using ATCA 
(Figure 1 of \citealt{Massardi.et.al2010}).
For comparison, we also plot the centers of the X--ray emission
to show that, in all of these cases, there is an offset between the SZ and X--ray peaks.
These panels illustrate that an observed nearly circular SZ distribution, with the
assumed SZ sensitivity and resolution, 
can not be used to identify a shock in a cluster, 
nor can the initial conditions be constrained using only SZ observations if we know that
there is a shock.
It is clear that we need a multi-frequency approach in this case.

Including the observed SZ and X--ray morphology of CL0152-1357 
(Figure 1 of \citealt{Massardi.et.al2010} and Figure 3 of \citealt{Maughet06ApJ640p219})
to constrain the initial conditions, 
we find that the best match with observations is provided by run 
RM0p7V48P15 with
masses of 2.1 \TMSUNFOUR\ and  0.7 \TMSUNFOUR, and $\Theta = 0$
(the collision in the pane of the sky), P = 150 kpc, and initial relative velocity
4800 \KMSEC\ 
(top panel in the right column of Figure~\ref{F:MULTI}).
We assumed a 12\ASEC\ X--ray resolution (XMM-Newton).
This high velocity seems to be needed to obtain the
large offset, about 45\ASEC\ between the SZ and X-ray peaks, 
and the observed SZ and X-ray morphology.
Allowing for some uncertainty in the relative pointing between the SZ and X-ray 
instruments, we find a good match with observations for run RM1V48p15 with 
2.1 \TMSUNFOUR\ and 1 \TMSUNFOUR, 
and the same impact parameter and relative velocity as the previous
run (middle panel in the right column of Figure~\ref{F:MULTI}).
This run with a rotation angle of $\Theta = 60$\Degree can also reproduces the observed 
SZ and X-ray morphology, but the SZ and X-ray offset is only about 20\ASEC.
Note, that the relative coordinate positions of the SZ and X--ray peaks differ
for these two cases, but that can not be observed using these two
wavelengths because we do not know the original orientation of the 
two clusters.
Even though the possible initial configurations are more restricted
by using X--ray and radio observations, 
we can conclude that, using morphology, 
these observations are not sufficient to confine the initial conditions of the collision.

We include now the mass surface density map of CL0152-1357 derived from gravitational
lensing observations in our analysis
assuming a resolution of 20\ASEC\ \citep{Jee_et05ApJ618p46}.
The importance of the lensing observations is that 
the distance between the two dark matter centers constrain the phase of
the collision, i.e. the time elapsed since the first core passage 
(subject to projection effects).
Having some constraints on the time elapsed since the first core passage and 
the X--ray and SZ morphology  
enable us to put more constraints on the initial conditions of the collision.

We are looking for an offset between the SZ and X--ray peaks of about
45\ASEC, and a distance of about 50\ASEC\
between the mass peaks, in projection.
For all projections perpendicular to the main plane of the collision
($\Theta=0$\ASEC), the distance between the mass peaks are around 
100\ASEC--120\ASEC, so we need to rotate the infalling cluster center towards the line of sight, 
thus we conclude that the rotation angle, $\Theta$, should be greater than zero.
We find that, with $\Theta$~=~60\Degree, our run RM1V48p15 
with P~=~150~kpc, M$_1$~=~2.1\TMSUNFOUR, M$_2$~=~1\TMSUNFOUR, 
and V~=~4800~\KMSEC, provides the best fit to all three observations.
The morphology of this model is close to the observed SZ and X--ray
morphologies \citep{Massardi.et.al2010,Maughet06ApJ640p219},
and the distance between the projected mass centers
is the same as the observed \citep{Jee_et05ApJ618p46}.
It is encouraging that these masses are similar to the
masses derived for these components by \citep{Jee_et05ApJ618p46}.
We were running simulations with larger masses (M$_1 > $ 2.1\TMSUNFOUR),
but the resulting X--ray morphologies did not match the observations.
The offset between the SZ and X--ray peaks is less than the 
observed 45\ASEC. This difference may be caused by absolute 
calibration for the positions between the ACT and \XMM.
However, it is also possible that the lensing observations have some
systematics. Gravitational lensing observations are very difficult, and 
the mass surface density reconstruction depends strongly on the 
assumed priors. Compare, for example, the results using different priors
for reconstruction of the mass surface density for RX J1347.5--1145
(see Figure 4 of \citealt{Mirandaet2008MNRAS385p511}).
It would be useful to have mass surface density reconstructions
using different methods to check for systematics in the analysis of  
CL0152-1357.

% % % % % % % % % % % % % % % % % % % % % % % % % % % % % % % % % % % % % % % % % % % %
%  
%  FIGURE 10
% 
% % % % % % % % % % % % % % % % % % % % % % % % % % % % % % % % % % % % % % % % % % % %
\begin{figure*}
\centerline{
\includegraphics[width=.95\textwidth]{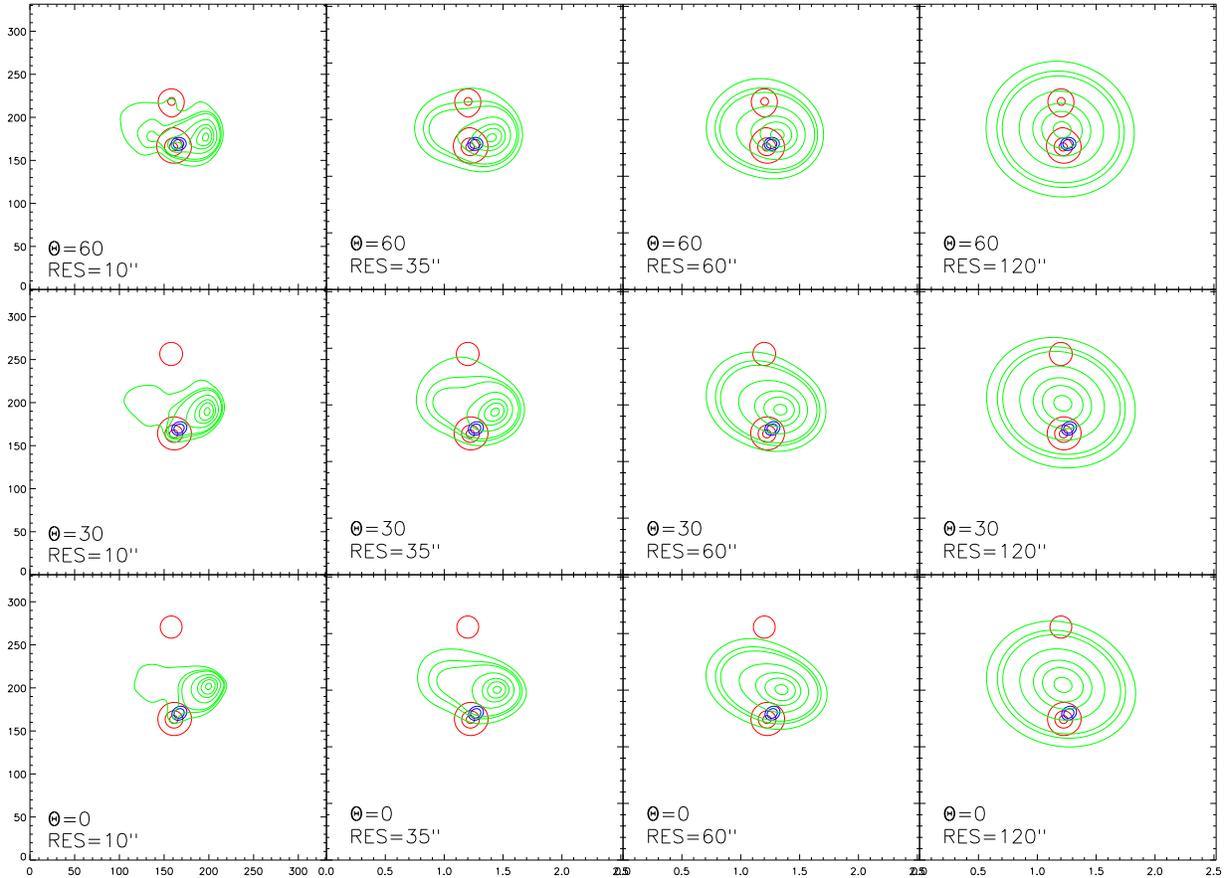}
}
\caption{
 Projected mass surface density, X-ray emission 
 and SZ contours red, blue and green contours) of 
 run RM1V48p15 (see Table~\ref{T:TABLE1}) as a function of
 rotation angle, $\Theta$, away from the plane of the collision for different
 resolutions, FWHM, of the SZ observations:
 FWHM = 10$\,\arcsec$, 35$\,\arcsec$, 60$\,\arcsec$, and 120$\,\arcsec$
 (left to right). The rotation angles are: 
$\Theta$ = 0\Degree, 30\Degree, 60\Degree\ away from main plane of the collision
 (from bottom to top).
 The images are $350\,\arcsec \times 350\,\arcsec$.
 \label{F:SZRESOL}
}
\end{figure*} % Figure~\ref{F:SZRESOL}

\smallskip
% % % % % % % % % % % % % % % % % % % % % % % % % % % % % % % % % % % % % % % % % % % %
\subsection{Tests for \LCDM}
\label{SS:TEST_LCDM}

Merging galaxy clusters can provide tests for our most successful cosmological 
scenario, the \LCDM\ models many ways. 
The features of mergers predicted by 
\LCDM\ models, for example, 
can be compared to those derived from observations
\citep{Foreet10ApJ725p598,MastBurk08MNRAS389p967}.

\cite{Foreet10ApJ725p598} derived the distribution of the offsets
between dark matter (DM) and gas centers of galaxy clusters 
drawn from one of the largest non-radiative, \LCDM\ cosmological simulations 
using SPH, usually referred to as the MareNostrum Universe. 
They found that the distribution of the 2D offsets between the DM and gas density peaks 
of the bullet cluster would have a probability of 1\%--2\% from simulations. 
Thus they concluded that it is possible to reproduce the observed offset between
DM and gas density peaks as large as observed in the bullet cluster 
assuming \LCDM\ cosmology.

The estimated initial relative velocities of clusters can also be used as a test for
\LCDM\ models.
\cite{MastBurk08MNRAS389p967} carried out galaxy cluster merger simulations 
and found that their best model requires a 
relative initial velocity of 3000 \KMSEC\ for the ``bullet'' cluster, 1E0657-56.
However, they concluded that the probability of finding such a high
relative velocity given their best model is less then 0.5\%.
We found a relative velocity of 4800 \KMSEC\ for our best model for
CL0152-1357.
The median relative velocity of a galaxy cluster with about the 
total mass of CL0152-1357 \citep{Jee_et05ApJ618p46}, 
$M_1$~=~5 \TMSUNFOUR, is about 1500 \KMSEC\
\citep{HayaWhit06MNRAS370pL38}.
The infall velocity of our best model is about three times the value of our
best fit model. Based on the results of \cite{HayaWhit06MNRAS370pL38}, 
this relative velocity is very unlikely in a \LCDM\ model.
These results suggest that there is a mismatch between 
the initial relative velocities derived from simple idealized cluster merging simulations and 
those derived from \LCDM\ simulations. There are two possibilities:
either our simple cluster models or our \LCDM\ models are missing some 
important physics. A detailed analysis of the origin of this mismatch is out of the scope 
of our paper, we are going to address this question in a future paper.

\smallskip
% % % % % % % % % % % % % % % % % % % % % % % % % % % % % % % % % % % % % % % % % % % %
\subsection{The Effect of SZ Resolution}
\label{SS:SZRESOL}

When using clusters to derive cosmological parameters we assume that they are
relaxed. The statistical methods use scaling relations derived based on relaxed clusters,
individual methods assume relaxed clusters to derive their physical parameters.
Both statistical and individual methods may be biased due to the fact that some
clusters are not relaxed. Statistical methods might have to be corrected for this effect,
when using individual methods, we can just simply exclude non-relaxed clusters.  
Some of these aspects of SZ resolution and the importance of high-resolution
SZ observations were discussed in 
\cite{Korngut.et.al2010,Mason.et.al2010ApJ716p739,Massardi.et.al2010}
for example.

In this paper we discuss some aspects of the identification of non-relaxed clusters
focusing on SZ observations.
The effect of non-relaxed clusters to determining cosmological parameters
using SZ observations was discussed in \cite{Wik.et.al2008ApJ680p17}.
The new generation of ground based SZ telescopes have a very large span in resolution, 
from about 10\ASEC\ to 3\AMIN.

We illustrate the importance of high resolution SZ observation in identifying merging clusters 
in Figure~\ref{F:SZRESOL}. As an example, we use the best model we found for CL0152-1357. 
In this Figure we show snapshots for this model after the first
core passage using different SZ resolution (green contours). 
We also plot the centers of projected mass (red contours) and X--ray peak (blue contours).
From this Figure we can see that, 
with high resolution, 10\ASEC--35\ASEC, SZ observations would be
sufficient to identify the merger based on its disturbed morphology.
Having also X--ray observations would help us because we can use
the offset between the SZ and X--ray peaks. 
In the case when the collision is in the plane of the sky ($\theta = 0$), this offset is about
50\ASEC\ with high SZ resolution, and about 40\ASEC\ for low, 1\AMIN -- 2\AMIN\
resolution. In this case the offset is observable assuming a better than 40\ASEC\ 
absolute pointing calibration for the SZ telescope, which is a reasonable assumption.
However, if the rotation angle is large, say 60\Degree, 
high resolution SZ instruments still measure an about 40\ASEC\ offset between the SZ and X--ray peaks, 
but with lower resolution, we obtain an only  10\ASEC\ offset, which is close or less then
the absolute pointing calibration of SZ instruments.
Therefore, we conclude that with an SZ resolution of about 10\ASEC\ and
35\ASEC, GBT/Mustang and ATCA have the sufficient resolution to identify 
this merger, but using lower resolution instruments of 1\AMIN or 2\AMIN, 
BOLOCAM and AMiBA, will not be able to identify the merger.
We could identify the merger in any case 
if we have not only SZ and X--ray measurements, but also
surface mass density maps from lensing optical/infrared (optical/IR) observations.

Our results suggest that if we want to identify mergers, we
should use either high resolution SZ observations, or 
multi--frequency observations.
In any case, galaxy cluster observations with 
low resolution SZ instruments with large field of view are still useful 
because they can be potentially used 
to constrain the large scale distribution of the intra-cluster gas 
(e.g., \citealt{Molnet10ApJ723p1272}).

\smallskip
% % % % % % % % % % % % % % % % % % % % % % % % % % % % % % % % % % % % % % % % % % % %
\section{Conclusion}
\label{S:Conclusion}

We have been studying galaxy cluster merging using self-consistent 
3D N-body/hydrodynamical simulations of idealized clusters.
We have found that significant offsets
results between the X--ray and SZ peaks (\simgreat $\,$40$\,$\ASEC)
if the relative velocities are equal or larger than about 4000 \KMSEC.
This offset can be observed using an SZ instrument with high
sensitivity and high angular resolution 
(at least about 10\ASEC--35\ASEC), in which case 
the merging cluster can be identified using SZ observations only
(Figure~\ref{F:SZRESOL}).
However, using multi-frequency observations seems to be a more practical way
to identify merging galaxy clusters.

We have applied numerical simulations to interpret
multi-frequency observations, radio (SZ), X--ray and optical/IR (lensing reconstruction) 
of CL0152-1357. 
We have found that the offsets between the SZ and X--ray peaks
provide important constraints on the initial parameters of the merging 
clusters, the offsets between the two peaks of the mass surface
densities provide constraints on the phase of the collision 
(i.e., the time elapsed after the first core passage), and that 
the SZ peak coincides with the peak in the pressure times the characteristic
length in the line of sight and not the pressure maximum (as it does for clusters in equilibrium).
The peak in the X--ray emission, as expected, coincides with the density maximum of the 
main cluster.
As a consequence, the morphology of the SZ signal, and therefore the offset 
between the SZ and X-ray peaks, change with viewing angle.

We conclude that analyzing the morphology of SZ, X--ray, and surface mass density 
images based on multi-frequency observations (radio, X--ray and optical/IR) 
enables us to put meaningful constraints on the masses of the colliding clusters,
the impact parameter and initial relative velocity of the collision.

\acknowledgements
We thank the referee for a thorough reading of the original manuscript and for
suggestions which helped to improve on the clarity of the presentation
of our results.
The code FLASH used in this work was in part developed by the DOE-supported 
ASC/Alliance Center for Astrophysical Thermonuclear Flashes at the University of Chicago. 
We thank the Theoretical Institute for Advanced Research in Astrophysics, Academia Sinica, 
for allowing us to use their high performance computer facility for our simulations.

% % % % % % % % % % % % % % % % % % % % % % % % % % % % % % % % % % % % % % % % % % % %
%
%                                      B I B L I O G R A P H Y
%
% % % % % % % % % % % % % % % % % % % % % % % % % % % % % % % % % % % % % % % % % % % %
\bibliographystyle{apj}

% % % % % % % % % % % % % % % % % % % % % % % % % % % % % % % % % % % % % % % % % % % %

\end{document}